# Dynamic model of firms competitive interaction on the market with taxation

Oleg Malafeyev[1], Eduard Abramyan[2], Shulga Andrey[3]

St.Petersburg State University, 7/9 Universitetskayanab., St.Petersburg, 199034 Russia

Abstract

In this article three models of firms interaction on the market are described. One of these models is described by using a differential equation and by Lotka-Volterra model, where the equation has a different form. Also, there are models of non-competing and competing firms. The article presents an algorithm for solving the interaction of competing firms in taxation and the calculation of a compromise point. Besides, the article presents a compromise between the interests of a state and an enterprise.

## 1 Introduction.

In this article dynamic model of competitive interaction of firms on the market with taxation is explored. Any enterprise such as any state has the same goal - to generate income. The company wants to get as much income as possible in the form of profit, and the goal of a state is to get as much income as possible through taxation of enterprises, institutions and organizations. It is not profitable for an enterprise to give part of its profit to the state, therefore, it reduces its income. But it is beneficial for the state to receive all the profit from the enterprise. For a normal existence, both of enterprise and states must seek a compromise. If the state takes all the profits from the enterprise, the enterprise will cease to exist. The funds that the company gives to the state in the form of tax, the state spends in favor of the same enterprise. The state establishes the rules in force on the market in this country, conducts anti-monopoly policy to maintain healthy competition on the market and protects the interests of domestic producers. All in all, the state guarantees the enterprise protection in case of violation of its legitimate interests by other market participants by the rules, which maintains competition on the market.

[1] malafeyevoa@mail.ru
[2] eduard-97@mail.ru
[3] head1.private@gmail.com

Many ideas that are used in this work are taken from [1]-[57]

## 2 Literature Review.

The following books are corresponded to the topic of a dynamic model of competitive interaction of firms on the market with taxation: the book Malafeyev's and Muraviev's books present some models, conflict situations and their resolution, which is related to this article. In the book of Christofides N., "Graph Theory", we can learn an algorithmic approach to competitive interaction. These list of books: "Game-theoretic model of two firms entering to the Cournot market", "Model of parallel computation", "Workshop on Computational Physics'' helps us to deepen in the issue. The book "Corruption dynamics model", Malafeyev O., Saifullina D., Ivaniukovich G., Marakhov V., Zaytseva I. is served as the theoretical basis for writing this article.

### 2.1 Informal statement of the problem.

Three models are presented in this article. In the first model, the two firms produce homogeneous goods and do not compete with each other. The second model generalizes the first, but in the second model, firms are competing with each other on the market. The third model is a model of competitive interaction between firms on the market and of the state, that influences their interaction through taxation. Any product has a double product. Products may vary in design, price and other parameters; by the way the manufacturer's firm (or the seller's firm) puts up for sale (positions) the goods; but the main purpose of the acquisition of goods is to be unchanged.
Let firms produce homogeneous goods. Then the rate of growth of the firm's capital depends on the volume of demand for the goods, produced by this firm, also, on the volume of demand for the alternate product, produced by a competing firm, and on the total volume of demand.
Further, the existence and uniqueness of the solution of this differential equation will be shown. We study the solution by analogy with Lotka-Volterra model.
$V_1$ is the capital of the first firm, and $V_2$ is the capital of the second firm $p_1$, $p_2$ - total demand for goods, these coefficients set the rate of growth of capital of firms.
$k_1$, $k_2$ - coefficients that simulate the process of saturation of demand, which goods produced by the first and the second firms.
We obtain equations that show the growth rates of capital of the first and the second firms.

$$V_1' = V_1(\rho_1 - \kappa_1(\kappa_1 V_1 + \kappa_2 V_2))$$
$$V_2' = V_2(\rho_2 - \kappa_2(\kappa_2 V_2 + \kappa_1 V_1)), \tag{1}$$

The first equation in the system (1) $k_1 V_1$ is a factor that simulates the saturation of demand at the expense of the goods produced by the first firm. $k_2 V_2$ is a term that simulates the saturation of demand by counteracting the competing firm. In the second equation the $k_2 V_2$ system is a factor that simulates the saturation of demand at the expense of the goods produced by the second company and $k_1 V_1$ is a term that simulates the saturation of demand at the expense of goods produced by a competing company.

$k_1 k_2$ - coefficients model how the demand is met with interchangeable goods at the expense of competitors. Equations (1) - Volterra equations, describe the competition model of two firms producing interchangeable goods.

## 3 Analysis of the differential equation solution of the model of interaction between two firms.

This system can be rewritten as (1 '):

$$\frac{dV_1}{dt} = V_1 \left( \rho_1 - \kappa_1 F(V_1, V_2) \right)$$
$$\frac{dV_2}{dt} = V_2 \left( \rho_2 - \kappa_2 F(V_1, V_2) \right)$$

We study the solution of this system with initial data $V_1^0$, $V_2^0$, positive for $t = t_0$

It can be shown that for any finite time interval $(t_0, T)$, there is a unique solution of two continuous functions between two positive numbers, which depends on the end of the interval T (i.e., $V_1$ and $V_2$ are bounded) .
Consider what happens with an unlimited increase in time.

$$\frac{d \log V_1}{dt} = \rho_1 - \kappa_1 F(V_1, V_2),$$
$$\frac{d \log V_2}{dt} = \rho_2 - \kappa_2 F(V_1, V_2),$$

We get

$$\kappa_2 \frac{d \log V_1}{dt} - \kappa_1 \frac{d \log V_2}{dt} = \rho_1 \kappa_2 - \rho_2 \kappa_1,$$

$$\frac{V_1^{\kappa_2}}{V_2^{\kappa_1}} = \frac{(V_1^0)^{\kappa_2}}{(V_2^0)^{\kappa_1}} e^{(\rho_1 \kappa_2 - \rho_2 \kappa_1)(t-t_0)}$$

then,

$$\frac{V_1^{\kappa_2}}{V_2^{\kappa_1}} = \frac{(V_1^0)^{\kappa_2}}{(V_2^0)^{\kappa_1}} e^{(\rho_1 \kappa_2 - \rho_2 \kappa_1)(t-t_0)} \qquad (2)$$

We can say that,

$$\rho_1 \kappa_2 - \rho_2 \kappa_1 = 0,$$

if

$$\rho_1 \kappa_2 - \rho_2 \kappa_1 > 0,$$

We get

$$\lim_{t \to \infty} \frac{V_1^{\kappa_2}}{V_2^{\kappa_1}} = +\infty$$

So, the value of the capital of the second company, in which $p_k$ is less important, the demand for the goods of this company is fully satisfied. It decreases and with time the second company loses all its capital, while the first company continues to exist. When a firm really loses all its capital, it is extremely rare situation. The owner of the company understands that the costs of the production and promotion of goods exceed the income from its sale, and take any steps to solve.

## 3.1 The case of non-competing firms.

If the second and the first firms exists on the market independently of each other, after large period of time the capital of the first firm obeys to the law.

$$\frac{dV_1}{dt} = V_1 \left( \rho_1 - \kappa_1 F(V_1, 0) \right),$$

Starting from the moment $t_1$, when $V_1$ takes the value $V_1'$.
Let's say $V_t^l$ is the root of the equation.

$$\rho_1 - \kappa_1 F(V_1, 0) = 0 \tag{3}$$

If $V_1' < V_2'$, the term (3), starting from the moment, will increase to the value $V_t^l$ following the law.

$$t - t_1 = \int_{V_1'}^{V_1} \frac{dV_1}{V_1(\rho_1 - \kappa_1 F(V_1, 0))}.$$

As was considered, that $F_{V_1}' \geq 0$, in neighbourhood $V_1^l$ for $V_1^1 < V_1^l$ turns out

$$F(V_1, 0) - F(V_1^l, 0) = (V_1 - V_1^l)\varphi(V_1)$$

by

$$F(V_1^l, 0) = \frac{\rho_1}{\kappa_1},$$

where φ is positive, it means that

$$\rho_1 - \kappa_1 F(V_1, 0) = \kappa_1(V_1^l - V_1)\varphi(V_1).$$

It can be concluded, that $V_1$ will never reach $V_1^l$, but any lower value can be achieved in a finite time.

## 3.2 The case of competing firms.

Consider a situation where two firms producing homogeneous goods compete with each other on the market
If we take for the function F as a first approximation

$$F(V_1, V_2) = \lambda_1 V_1 + \lambda_2 V_2,$$

where $\lambda_1 > 0, \lambda_2 > 0, V_1$ starting at time $t_1$, it changes according to the law

$$\frac{dV_1}{dt} = V_1(\rho_1 - \kappa_1\lambda_1 V_1)$$

If $V_1 = V_1'$, we can have

$$t - t_1 = \int_{V_1'}^{V_1} \frac{dV_1}{V_1(\rho_1 - \kappa_1\lambda_1 V_1)}$$

As

$$\frac{1}{V_1(\rho_1 - \kappa_1\lambda_1 V_1)} = \frac{1}{\rho_1}\left[\frac{1}{V_1} + \frac{\kappa_1\lambda_1}{\rho_1 - \kappa_1\lambda_1 V_1}\right],$$

We obtain

$$t - t_1 = \frac{1}{\rho_1}\left[\int_{V_1'}^{V_1} \frac{dV_1}{V_1} + \int_{V_1'}^{V_1} \frac{\kappa_1\lambda_1}{\rho_1 - \kappa_1\lambda_1 V_1} dV_1\right] =$$

$$= \frac{1}{\rho_1}\left[(\log V_1)\big|_{V_1'}^{V_1} + (-\log|\rho_1 - \kappa_1\lambda_1 V_1|)\big|_{V_1'}^{V_1}\right] =$$

$$= \frac{1}{\rho_1}\left(\log \frac{V_1}{V_1'} - \log\left|\frac{\rho_1 - \kappa_1\lambda_1 V_1}{\rho_1 - \kappa_1\lambda_1 V_1'}\right|\right) =$$

$$= \frac{1}{\rho_1}\log\left|\frac{V_1}{\rho_1 - \kappa_1\lambda_1 V_1} * \frac{\rho_1 - \kappa_1\lambda_1 V_1'}{V_1'}\right|$$

From here,

$$\left|\frac{V_1}{\rho_1 - \kappa_1\lambda_1 V_1}\right| = \left|\frac{V_1'}{\rho_1 - \kappa_1\lambda_1 V_1'}\right| e^{\rho_1(t-t_1)}$$

$$\frac{V_1}{\rho_1 - \kappa_1\lambda_1 V_1} = \frac{V_1'}{\rho_1 - \kappa_1\lambda_1 V_1'} e^{\rho_1(t-t_1)}$$

Solving this equation, we obtain this.

$$D = \frac{V_1'}{\rho_1 - \kappa_1\lambda_1 V_1'}$$

If the capital of the first firm has a finite redistribution, different from zero, then the second firm ceases to exist.

This does not always mean the bankruptcy of the company. In nowadays companies have not one, but companies have several areas of activity.

The management of the company, which has only one line of business, may not bring the case to bankruptcy, but simply makes a decision to liquidate the company, the line of business, or change the product being produced for the better.

A company with several fields of activity has the same choice - to close unprofitable production, or to change a product, or to organize a more successful promotion of this product on the market; but it is much easier for a large company to implement any of these measures, since it is financially more stable and has the money that it can invest in the above measures. Based on the above, we calculate the total income of firms competing with each other on the market for a finite period of time without taxation:

$$\int_0^T (V_1 + V_2)dt = \int_0^T \frac{D_1\rho_1 e^{\rho_1(t-t_1)}}{1 + \kappa_1^2 D e^{\rho_1(t-t_1)}}dt +$$

$$+ \int_0^T \frac{D_2\rho_2 e^{\rho_2(t-t_2)}}{1 + \kappa_2^2 D e^{\rho_2(t-t_2)}}dt = \frac{\ln(e^{t_1\rho_1} + D_1\kappa_1^2 \rho_1 e^{\rho_1 T})}{\kappa_1^2} -$$

$$- \frac{\ln(e^{t_1\rho_1} + D_1\kappa_1^2)}{\kappa_1^2} + \frac{\ln(e^{t_2\rho_2} + D_2\kappa_2^2 \rho_2 e^{\rho_2 T})}{\kappa_2^2} -$$

$$- \frac{\ln(e^{t_2\rho_2} + D_2\kappa_2^2)}{\kappa_2^2},$$

Where,

$$D_1 = \frac{V_1^0}{\rho_1 - \kappa_1^2 V_1^0},$$

$$D_2 = \frac{V_2^0}{\rho_2 - \kappa_2^2 V_2^0},$$

$V_1^0$ – first company capital value, $V_2^0$ - second company capital value.

## 4 Study based on Lotka Volterra model.

When we neglect the case in which the firm itself saturates the market with its goods, the demand for goods produced by the firm is reduced only by the products of a competing firm. Then equation (1) has the form:

$$V_1' = V_1(\rho_1 - \kappa_2 V_2))$$

$$V_2' = V_2(\rho_2 - \kappa_1 V_1)), \qquad (4)$$

So, from system (4), we can write out three solutions to this system:

1) $V_1(t) = V_2(t) = 0$
2) $V_1(t) = 0, V_2(t) = V_2(0)e^{\rho_2 t}$ ( $V_2(0) > 0$)
3) $V_2(t) = 0, V_1(t) = V_1(0)e^{\rho_1 t}$ ( $V_1(0) > 0$)

If the values of $V_1$ or $V_2$ are equal to zero at any given time, then they remain equal to zero in the future. This means that if the capital of the first firm or the capital of the second firm is zero, then there will be no change. The first solution corresponds to the origin, the second solution corresponds to the half-axis $V_2 > 0$, the third solution corresponds to the half-axis $V_1 > 0$, and they form the boundary of the positive orthant

$$R_+^2 = \{(V_1, V_2) \in R^2 : V_1 \geq 0, V_2 \geq 0\}$$

By virtue of the non-negativity of the quantities $V_1$, $V_2$, system (4) can be considered on the set $R_+^2$. This set is invariant in the sense that any solution starting with $R_+^2$ does not leave it with the passage of time. As already mentioned, the boundary $bdR_+^2$ of the set $R_+^2$ is an invariant set. Since the trajectories do not intersect each other, the inner part of the set

$$int R_+^2 = \{(x, y) \in R^2 : x > 0, y > 0\}$$

is an invariant set too.

By virtue of the type of system (4), there exists a unique equilibrium position belonging to the notation F $(V_1, V_2)$ must satisfy the set $int R_+^2$. next it is clear that the position are:

$$\bar{V}_1(\rho_1 - \rho_2 \bar{V}_2) \qquad \bar{V}_2(\rho_2 + \kappa_1 \bar{V}_1)$$

Multiplying the first equation of system (4) by the term $p_2 - k_1 V_1$, the second equation by the term $p_2 - k_2 V_2$ and adding them, we get

$$\frac{d}{dt}[\rho_2 \ln V_1 - \kappa_1 V_1 + \rho_1 \ln V_2 - \kappa_2 V_2] = 0$$

then,

$$H(V_1) = \bar{V}_1 \ln V_1 - V_1, \qquad G(V_2) = \bar{V}_2 \ln V_2 - V_2$$

and

$$X(V_1, V_2) = dH(V_1) + bG(V_2),$$

Equation (5) can be written as:

$$\frac{d}{dt} X(V_1, V_2) = 0$$

The function X defined by system (4) and remains constant on solutions of constant motion. As the functions H and G are

$$\frac{dH}{dV_1} = \frac{\bar{V}_1}{V_1} - 1, \quad \frac{d^2H}{dV_1^2} = -\frac{\bar{V}_1}{V_1^2} < 0;$$

$$\frac{dG}{dV_2} = \frac{\bar{V}_2}{V_2} - 1, \quad \frac{d^2G}{dV_1^2} = -\frac{\bar{V}_2}{V_2^2} < 0$$

Then the function X $(V_1, V_2)$ reaches its maximum at the point F = (x, y). Level lines by ty curves $(V_1, V_2) \in$ around the point $intR_2^+$ Solutions: X $(V_1, V_2)$ = should const be closed - remain on the level lines and return to the starting point. The solutions of system (4) are periodic. The process of interaction between two competing firms on the market described by the system of differential equations (4) is periodic in nature. The amount of capital of the first and the second firms are fluctuate, and the amplitude and frequency of these fluctuations are determined by the initial conditions. However, the time-averaged values of $V_1$, $V_2$, are denoting the volume of stock and the number of enterprises, is constant and equal to the values to the equilibrium position:

$$\frac{1}{T}\int_0^T V_1(t)dt = \bar{V}_1, \quad \frac{1}{T}\int_0^T V_2(t)dt = \bar{V}_2$$

where T is the period of oscillation. Integrating the term

$$\frac{d}{dt}(\ln V_1) = \frac{\dot{V}_1}{V_1} = \rho_1 - \kappa_2 V_2$$

then,

$$\int_0^T \frac{d}{dt}\ln V_1(t)dt = \int_0^T (\rho_1 - \kappa_1 V_2(t))dt$$

or

$$\ln V_1(T) - \ln V_1(0) = \rho_1 T - \kappa_2 \int_0^T V_2(t)dt.$$

If $V_1(T) = V_1(0)$, it becomes

$$\frac{1}{T}\int_0^T V_2(t)dt = \frac{\rho_1}{\kappa_2} = \bar{V}_2.$$

# 5 Case of competitive interaction firms under taxation.

There are cases when the state taxes corporate incomes. Let $U_1$ be the value of tax levied by the state from the first firm. $U_2$ is the amount of tax paid by the second firm to the state. Then the corresponding equations of dynamics looks as follows.

$$V_1' = V_1(\rho_1 - \kappa_1^2 V_1 - \kappa_1 \kappa_2 V_2) - U_1$$
$$V_2' = V_2(\rho_2 - \kappa_2^2 V_2 - \kappa_1 \kappa_2 V_1) - U_2$$

(6)

If the state taxes the income of the company and the tax is a percentage of the income, we have to fix the share of capital that is taxable per unit of time. U = xV, where $x \in (0,1)$

Then we obtain differential equations that show the growth rate of capital of competing firms in taxation:

$$V_1' = V_1(\rho_1 - \kappa_1^2 V_1 - \kappa_1 \kappa_2 V_2) - xV_1$$
$$V_2' = V_2(\rho_2 - \kappa_2^2 V_2 - \kappa_1 \kappa_2 V_1) - xV_2$$

(7)

The problem arises by the time T to calculate the total income consisting of the capital of the first firm, the capital of the second firm and the state's income.

$$\int_0^T (V_1 + V_2 + U(V_1) + U(V_2))dt$$

Solving the system of equations (7). We can get

$$V_1 = \frac{\rho_1 - x}{\kappa_1^2 + e^{t(x-\rho_1)} D_1 \rho_1 - e^{t(x-\rho_1)} D_1 x}$$

$$V_2 = \frac{\rho_2 - x}{\kappa_2^2 + e^{t(x-\rho_2)} D_2 \rho_1 - e^{t(x-\rho_2)} D_2 x},$$

where,

$$D_1 = \frac{V_1^0}{\rho_1 - \kappa_1^2 V_1^0},$$

$$D_2 = \frac{V_2^0}{\rho_2 - \kappa_2^2 V_2^0}$$

Let's calculate the total income for a finite period of time.

$$\int_0^T (V_1 + V_2 + U_1 + U_2)dt = \int_0^T (V_1 + V_2 + xV_1 + xV_2)dt =$$

$$= \left(-\frac{\rho_1 \ln(\kappa_1^2 + e^{t(x-\rho_1)}D_1\rho_1 - e^{t(x-\rho_1)}D_1x)}{\kappa_1^2(x-\rho_1)}\right)\Bigg|_0^T +$$

$$+ \left(\frac{\rho_1(x-\rho_1)t}{\kappa_1^2(x-\rho_1)}\right)\Bigg|_0^T + \left(\frac{x\ln(\kappa_1^2 + e^{t(x-\rho_1)}D_1\rho_1 - e^{t(x-\rho_1)}D_1x)}{\kappa_1^2(x-\rho_1)}\right)\Bigg|_0^T -$$

$$- \left(\frac{x(x-\rho_1)t}{\kappa_1^2(x-\rho_1)}\right)\Bigg|_0^T - \left(\frac{\rho_1 x \ln(\kappa_1^2 + e^{t(x-\rho_1)}D_1\rho_1 - e^{t(x-\rho_1)}D_1x)}{\kappa_1^2(x-\rho_1)}\right)\Bigg|_0^T +$$

$$+ \left(\frac{\rho_1(x-\rho_1)xt}{\kappa_1^2(x-\rho_1)}\right)\Bigg|_0^T +$$

$$+ \left(\frac{x^2 \ln(\kappa_1^2 + e^{t(x-\rho_1)}D_1\rho_1 - e^{t(x-\rho_1)}D_1x)}{\kappa_1^2(x-\rho_1)}\right)\Bigg|_0^T - \left(\frac{x^2(x-\rho_1)t}{\kappa_1^2(x-\rho_1)}\right)\Bigg|_0^T -$$

$$- \left(\frac{\rho_2 \ln(\kappa_2^2 + e^{t(x-\rho_2)}D_2\rho_2 - e^{t(x-\rho_2)}D_2x)}{\kappa_2^2(x-\rho_2)}\right)\Bigg|_0^T +$$

$$+ \left(\frac{\rho_2(x-\rho_2)t}{\kappa_2^2(x-\rho_2)}\right)\Bigg|_0^T +$$

$$+ \left(\frac{x\ln(\kappa_2^2 + e^{t(x-\rho_2)}D_2\rho_2 - e^{t(x-\rho_2)}D_2x)}{\kappa_2^2(x-\rho_2)}\right)\Bigg|_0^T - \left(\frac{x(x-\rho_2)t}{\kappa_2^2(x-\rho_1)}\right)\Bigg|_0^T -$$

$$- \left(\frac{\rho_2 x \ln(\kappa_2^2 + e^{t(x-\rho_2)}D_2\rho_2 - e^{t(x-\rho_2)}D_2x)}{\kappa_2^2(x-\rho_2)}\right)\Bigg|_0^T +$$

$$+ \left(\frac{x^2 \ln(\kappa_2^2 + e^{t(x-\rho_2)}D_2\rho_2 - e^{t(x-\rho_2)}D_2x)}{\kappa_2^2(x-\rho_2)}\right)\Bigg|_0^T - \left(\frac{x^2(x-\rho_2)t}{\kappa_2^2(x-\rho_2)}\right)\Bigg|_0^T =$$

$$= \frac{(x+1)}{\kappa_1^2}(\ln(\kappa_1^2 + e^{T(x-\rho_1)}D_1\rho_1 - e^{T(x-\rho_1)}D_1x)-$$

$$-T(x-\rho_1) + \ln(\kappa_1^2 + D_1\rho_1 - D_1x))+$$

$$+ \frac{(x+1)}{\kappa_2^2}(\ln(\kappa_2^2 + e^{T(x-\rho_2)}D_2\rho_2 - e^{T(x-\rho_2)}D_2x)-$$

$$-T(x-\rho_2) + \ln(\kappa_2^2 + D_2\rho_2 - D_2x))$$

Our task is to find a compromise between competing firms and the state.
Let's describe the algorithm for finding it.

## 5.1 Solution Algorithm

We can describe the algorithm for finding a compromise point.
   STEP 1: Calculate the income of the state and each company for a finite period of time.
   STEP 2: For the state and each company we find the maximum income Ci.
   STEP 3: Find for the state and each firm a deviation from the maximum (Ci) for the remaining income functions.
   STEP 4: From the deviations found for the state and each company, we choose the maximum.
STEP 5: Choose the minimum of these maximum deviations, then the tax rate in which this minimum is reached and is a compromise for the state and competing firms.

## 5.2 Calculation of a compromise point.

Calculate the income of the state and each company for a finite period of time.

If,

$$h_1 = \int_0^T V_1 dt = \frac{1}{\kappa_1^2}(-\ln(\kappa_1^2 + e^{T(x-\rho_1)}D_1\rho_1 - e^{T(x-\rho_1)}D_1 x) +$$

$$+ T(x - \rho_1) + \ln(\kappa_1^2 + D_1\rho_1 - D_1 x))$$

income of the first company for a finite period of time.

$$h_2 = \int_0^T V_2 dt = \frac{1}{\kappa_2^2}(-\ln(\kappa_2^2 + e^{T(x-\rho_2)}D_2\rho_2 - e^{T(x-\rho_2)}D_2 x) +$$

$$+ T(x - \rho_2) + \ln(\kappa_2^2 + D_2\rho_2 - D_2 x))$$

income of the second company for a finite period of time.

$$h_3 = \int_0^T x V_1 + x V_2 dt = \frac{x}{\kappa_1^2}(-\ln(\kappa_1^2 + e^{T(x-\rho_1)}D_1\rho_1 - e^{T(x-\rho_1)}D_1 x) +$$

$$+ T(x-\rho_1) + \ln(\kappa_1^2 + D_1\rho_1 - D_1 x)) + \frac{x}{\kappa_2^2}(-\ln(\kappa_2^2 + e^{T(x-\rho_2)}D_2\rho_2 - e^{T(x-\rho_2)}D_2 x) +$$

$$+ T(x - \rho_2) + \ln(\kappa_2^2 + D_2\rho_2 - D_2 x))$$

state income derived from the collection of taxes for a finite period of time. For the state and each company we find the maximum income i. Firms gets the maximum profit with minimum taxation, i.e. when their income is not taxed, it happens at x = 0

$$C_1 = \max \int_0^T V_1 dt = \frac{-\ln(\kappa_1^2 + e^{-T\rho_1}D_1\rho_1) - T\rho_1 + \ln(\kappa_1^2 + D_1\rho_1)}{\kappa_1^2}$$

$$C_2 = \max \int_0^T V_2 dt = \frac{-\ln(\kappa_2^2 + e^{-T\rho_2}D_1\rho_2) - T\rho_2 + \ln(\kappa_2^2 + D_2\rho_2)}{\kappa_2^2}$$

The state receives the maximum profit at maximum taxation, i.e. with x = 1

$$C_3 = \max \int_0^T xV_1 + xV_2 dt = \frac{1}{\kappa_1^2}(-\ln(\kappa_1^2 + e^{T(1-\rho_1)}D_1\rho_1 - e^{T(1-\rho_1)}D_1) +$$

$$+T(1-\rho_1) + \ln(\kappa_1^2 + D_1\rho_1 - D_1)) + \frac{1}{\kappa_2^2}(-\ln(\kappa_2^2 + e^{T(1-\rho_2)}D_2\rho_2 - e^{T(1-\rho_2)}D_2) +$$

$$+T(1-\rho_2) + \ln(\kappa_2^2 + D_2\rho_2 - D_2))$$

Let's find the deviation from the maximum of the other income functions for the first company:

$$h_2 - C_2 = \frac{1}{\kappa_2^2}(-\ln(\kappa_2^2 + e^{T(x-\rho_2)}D_2\rho_2 - e^{T(x-\rho_1)}D_2x) +$$

$$+Tx + \ln(\kappa_2^2 + D_2\rho_2 - D_2x) + \ln(\kappa_2^2 + e^{-T\rho_2}D_2\rho_2) - \ln(\kappa_2^2 + D_2\rho_2))$$

Let's find the deviation from the maximum of the other income functions for the second company:

$$h_2 - C_2 = \frac{1}{\kappa_2^2}(-\ln(\kappa_2^2 + e^{T(x-\rho_2)}D_2\rho_2 - e^{T(x-\rho_1)}D_2x) +$$

$$+Tx + \ln(\kappa_2^2 + D_2\rho_2 - D_2x) + \ln(\kappa_2^2 + e^{-T\rho_2}D_2\rho_2) - \ln(\kappa_2^2 + D_2\rho_2))$$

Then, we find the deviation from the maximum of the other functions of income for the state:

$$h_3 - C_3 = \frac{x}{\kappa_1^2}(-\ln(\kappa_1^2 + e^{T(x-\rho_1)}D_1\rho_1 - e^{T(x-\rho_1)}D_1 x)+$$

$$+T(x-\rho_1)+\ln(\kappa_1^2+D_1\rho_1-D_1 x))+\frac{x}{\kappa_2^2}(-\ln(\kappa_2^2+e^{T(x-\rho_2)}D_2\rho_2-e^{T(x-\rho_2)}D_2 x)+$$

$$+T(x-\rho_2)+\ln(\kappa_2^2+D_2\rho_2-D_2 x))-\frac{1}{\kappa_1^2}(-\ln(\kappa_1^2+e^{T(1-\rho_1)}D_1\rho_1-e^{T(1-\rho_1)}D_1)+$$

$$+T(1-\rho_1)+\ln(\kappa_1^2+D_1\rho_1-D_1))+\frac{1}{\kappa_2^2}(-\ln(\kappa_2^2+e^{T(1-\rho_2)}D_2\rho_2-e^{T(1-\rho_2)}D_2)+$$

$$+T(1-\rho_2)+\ln(\kappa_2^2+D_2\rho_2-D_2))$$

Let's find the maximum deviation for each agent:

$$\max(C_1 - h_1) = \min(h_1 - C_1)$$
$$\max(C_2 - h_2) = \min(h_2 - C_2)$$
$$\max(C_3 - h_3) = \min(h_3 - C_3)$$

Our goal is to find a compromise point, therefore, we need to choose the minimum of these maximum deviations, and the point at which this is achieved is a compromise for the state and competing firms.

$$\min\max(C_i - h_i) = \max\min(h_i - C_i)$$

[6] Jackson M.O., Wolinsky A. A Strategic Model of Social and Economics Networks. //J. Econom. Theory. 1996. ¹ 71 P.44–74.

[7] Malafeev O.A. Controlled Systems of Conflict. // Sankt-Peterburg. Izdatelstvo SPbGU, 2000. 276 pp.

[8] Malafeev O.A., Boitsov D.S., Redinskikh N.D., Neverova E.G. Compromise solution and equilibrium in multi-agent control models of social networks with corruption. // Young scientist. 2014. ¹10 (69). pp. 14–17.

[9] Malafeev O.A., Sosnina V.V. Model of controlled process of cooperative 3-agent interaction.// Problems of mechanics and control. Nonlinear dynamic systems: intercollegiate collection of scientific papers. Perm State University./. Perm', 2009. V.39. pp.131–144.

[10] Anna Nagurney and Dong Li A Supply Chain Network Game Theory Model with Product Differentiation, Outsourcing of Production and Distribution, and Quality and Price Competition. // Analysis of Operations Research (2015), 228(1), pp. 479–503.

[11] Parfenov A.P., Malafeev O.A. Equilibrium and compromise solution in network models of muli-agent interaction. // Problems of mechanics and control. Nonlinear dynamic systems: intercollegiate collection of scientific papers. Perm State University. /Perm', 2007.V.39. pp.154–167.

[12] Petrosyan L.A., Sedakov A.A. One-Way Flow Two-Stage NetworkGames. // Bulletin of St. Petersburg University. Series 10. Applied Mathematics, Computer Science, Control Processes. No. 4. 2014. pp.72–81.
[13] Anu Thomas, Mohan Krishnamoorthy, Jayendran Venkateswaran Gaurav Singh Decentralised decision-making in a multi-party supply chain. // International Journal of Production Research, 54:2, 2016. pp.405–425. 15

[14] O.A. Malafeyev, Dynamical processes of conflict, St. Petersburg State University. Saint-Petersburg, 1993, 95.

[15] F. L.Chernousko, N.N.Bolotnik, V.G.Gradetskiy, Manipulation robots: dynamics, control, optimization. Science, 1989, 368. (in Russian) 10

[16] G.V. Alferov, O.A. Malafeyev, A.S. Maltseva, Game-theoretic model of inspection by anti-corruption group, AIP Conference Proceedings, (2015),1648, http://dx.doi.org/10.1063/1.4912668.

[17] V.N. Kolokoltsov, O.A. Malafeyev, Mean-Field-Game Model of Corruption, Dynamic Games and Applications, (2015), http://dx.doi.org/10.1007/s13235-015-0175-x.

[18] O.A. Malafeyev, N.D. Redinskikh, G.V. Alferov, Electric circuits analogies in economics modeling: Corruption networks, Proceedings of 2nd
15